\documentclass[aps,pra,twocolumn,groupedaddress,showpacs]{revtex4}
\usepackage{epsfig}

\usepackage{bbm}
\usepackage{graphicx}
\usepackage{color}
\begin{document}

\definecolor{drk}{rgb}{0,0,0}
\definecolor{frm}{rgb}{0,0,1}

\color{drk}

\title{Superfluid and Mott Insulator phases of one-dimensional Bose-Fermi mixtures}
\author{A.~Zujev}
\affiliation{Physics Department, University of California, Davis, California 95616, USA}
\author{A.~Baldwin}
\affiliation{Physics Department, University of California, Davis, California 95616, USA}
\author{R.~T.~Scalettar}
\affiliation{Physics Department, University of California, Davis, California 95616, USA}
\author{V.~G.~Rousseau}
\affiliation{Lorentz Institute, Leiden University, P. O. Box 9506, 2300 RA  Leiden, The Netherlands}
\author{P.~J.~H.~Denteneer}
\affiliation{Lorentz Institute, Leiden University, P. O. Box 9506, 2300 RA  Leiden, The Netherlands}
\author{M.~Rigol}
\affiliation{Physics Department, University of California, Santa Cruz, California 95064, USA}
\affiliation{Department of Physics, Georgetown University, Washington, District of Columbia 20057, USA}

\begin{abstract}
We study the ground state phases of Bose-Fermi mixtures in
one-dimensional optical lattices with quantum Monte Carlo simulations
using the Canonical Worm algorithm. Depending on the filling of bosons
and fermions, and the on-site intra- and inter-species interaction,
different kinds of incompressible and superfluid phases appear.  On the
compressible side, correlations between bosons and fermions can lead to
a distinctive behavior of the bosonic superfluid density and the
fermionic stiffness, as well as of the equal-time Green functions, which
allow one to identify regions where the two species exhibit
anticorrelated flow.  We present here complete phase diagrams for these
systems at different fillings and as a function of the interaction
parameters.
\end{abstract}

\maketitle

\section{Introduction}

The experimental realization of strongly correlated systems with
ultracold gases loaded in optical lattices \cite{greiner02} has
generated tremendous excitement during recent years. Initially thought
of as a way to simulate condensed matter model Hamiltonians, like the
Bose-Hubbard Hamiltonian \cite{jaksch98}, loading atoms on optical
lattices has enabled the creation of quantum systems that are unexpected
in the condensed matter context. Among these systems the realization of
Bose-Fermi mixtures in optical lattices
\cite{ott04,gunter06,ospelkaus06}, where the inter- and intra-species
interactions can be tuned to be attractive or repulsive
\cite{zaccanti06}, is a remarkable example of the scope of realizable
models.

Theoretical studies of Bose-Fermi mixtures in one-dimensional lattices
have been done for homogeneous
\cite{cazalilla03,lewenstein04,mathey04,imambekov06,pollet06,sengupta07,hebert07,mering08}
and trapped \cite{albus03,cramer04,pollet06a} systems. Several
approaches have been used: Gutzwiller mean-field theory \cite{albus03},
strong coupling expansions \cite{lewenstein04,mering08}, bosonization
\cite{cazalilla03,mathey04} and exact analytical \cite{imambekov06} and
numerical \cite{pollet06,sengupta07,hebert07,pollet06a,mering08} studies.
Recently, a mixture of bosonic atoms and molecules on a lattice was
studied numerically \cite{rousseau01}.  The landscape of phases
encountered is expansive, and includes Mott insulators, spin and charge
density waves, a variety of superfluids, phase separation, and Wigner
crystals. However, the phase diagram in the chemical
potential-interaction strength plane has not yet been reported.

It is our goal in this paper to present a study of repulsive Bose-Fermi
mixtures in one-dimensional lattices that generalizes previous studies,
which focused on specific special densities, to more general filling.
After mapping the phase diagram we will explore different sections in
greater detail.  Since the particular case in which the lattice is half
filled with bosons and half filled with fermions has been carefully studied 
in Ref.~\cite{pollet06}, we will instead concentrate here
on two cases: (i) when the number of bosons is commensurate with the
lattice size but the number of fermions is not, and (ii) when the sum of
both species is commensurate with the lattice size but the number of
bosons and fermions are different. Some of the phases present in these
cases have been identified by Sengupta and Pryadko in their grand
canonical study in Ref.~\cite{sengupta07} and by H\'ebert {\it et
al.}\ in the canonical study recently presented in
Ref.~\cite{hebert07}.

The Hamiltonian of Bose-Fermi mixtures in one dimension can be written as
\begin{eqnarray}
\hat H = &-&t_B \sum_{l} (\, b^\dagger_{l+1} b^{\phantom\dagger}_l + b^\dagger_l b^{\phantom\dagger}_{l+1}\, )
\nonumber \\
&-&t_F \sum_{l} (\, f^\dagger_{l+1} f^{\phantom\dagger}_l + f^\dagger_l f^{\phantom\dagger}_{l+1}\, )
\nonumber \\
&+& U_{BB} \sum_l \, \hat n^B_l (\hat n^B_l -1) + U_{BF} \sum_l \, \hat n^B_l \hat n^F_l
\end{eqnarray}
where $b_{l}^{\dagger}(b^{\phantom\dagger}_l)$ are the boson creation
(destruction) operators on site $l$ of the one-dimensional lattice with
$L$ sites.  Similarly, $f_{l}^{\dagger}(f^{\phantom\dagger}_l)$ are the
creation (destruction) operators on site $l$ for spinless fermions on
the same lattice. For these creation and destruction operators $\hat
n^{B,F}_l$ are the associated number operators.  The bosonic and
fermionic hopping parameters are denoted by $t_B$ and $t_F$
respectively, and the on-site boson-boson and boson-fermion interactions
by $U_{BB}$ and $U_{BF}$. In this paper we will consider the case
$t_{B}=t_{F}=1$ (i.e.~when the boson and fermion hopping integrals are
equal) and choose $t_B=1$ to set the scale of energy.

It is useful to begin a discussion of the phase diagram with an analysis
of the zero hopping limit ($t_B=t_F=0$) similar to the one done by
Fisher {\it et al.}\ in Ref.~\cite{fisher89} for the purely bosonic
case. Consider a particular fermion occupation of one fourth of the
lattice sites, $N_F=L/4$ fixed.  Bosons can be added up to $N_B=3L/4$
without sitting on a site which is already occupied by either a boson or
a fermion.  Therefore the associated chemical potential $\mu$ is small.  What
happens when $N_B$ exceeds $3L/4$ depends on the relative strength of
$U_{BB}$ and $U_{BF}$.

If $U_{BF}$ is less than $2 U_{BB}$ then the extra bosons sit atop of
the fermions and $\mu$ jumps by $U_{BF}$.  The chemical potential stays
at this elevated value of $U_{BF}$ until all the sites with fermions
also have a boson.  At that point additional bosons start going onto
sites with a boson already, and $\mu$ jumps to $2 U_{BB}$.  Thus in
general there are incompressible phases where the boson chemical
potential jumps both at commensurate $\rho_B=1, \, 2, \, 3, ...$ (as for
the pure boson-Hubbard model) and also at $\rho_B= 1 - \rho_F, \,\, 2 -
\rho_F, \,\, 3 - \rho_F, \cdots $.  For $U_{BF}$ greater than $2 U_{BB}$
and less than $6 U_{BB}$ the incompressible phases still start at
$\rho_B= 1 - \rho_F$ but the following potential jumps are shifted up by
$1 - \rho_F$.  Turning on the hoppings $t_B, t_F$ introduces quantum
fluctuations which will ultimately destroy these Mott plateaus and
introduce new, intricate phases.

\section{Canonical Worm Algorithm}

We perform Quantum Monte Carlo simulations (QMC) using a recently
proposed Canonical Worm algorithm \cite{vanhoucke06,rombouts06}.  This
approach makes use of global moves to update the configurations, samples
the winding number, and gives access to the measurement of $n$-body
Green functions. It also has the useful property of working in the
\textit{canonical} ensemble. This is particularly important for the
present application since working with two species of particles leads to
two different chemical potentials in the grand canonical ensemble.
These prove difficult to adjust such that the precise, desired fillings
are achieved. In our canonical simulations the Bose and Fermi
occupations are exactly specified and the chemical potentials $\mu_B$
and $\mu_F$ are instead computed \cite{batrouni90} via appropriate
numerical derivatives of the resultant ground state energy (e.g.\ $\mu_B
= E_0(N_B+1) -E_0(N_B)$).

The Canonical Worm algorithm is a variation of the Prokof'ev {\it et~al.}\
grand-canonical worm algorithm \cite{prokofev98}.  Within the Canonical
Worm approach one starts by writing the Hamiltonian as $\hat H=\hat
V-\hat T$, where $\hat T$ is comprised of the non-diagonal terms and is
by necessity positive definite. The partition function $\mathcal
Z=\textrm{Tr}e^{-\beta\hat H}$ takes the form 
\begin{eqnarray}
\label{Eq-Worm-1} \mathcal Z &=& \textrm{Tr }e^{-\beta\hat V}
\textbf{T}_\tau e^{\int_0^\beta \hat T(\tau)d\tau} \\ \label{Eq-Worm-2}
&=& \textrm{Tr }e^{-\beta\hat V} \sum_n
\int_{0<\tau_1<\cdots<\tau_n<\beta}
\!\!\!\!\!\!\!\!\!\!\!\!\!\!\!\!\!\!\!\!\!\!\!\!\!\!\!\!\!\!\!\! \hat
T(\tau_n)\cdots\hat T(\tau_1)d\tau_1\cdots d\tau_n 
\end{eqnarray} 
where
$\hat T(\tau)=e^{\tau\hat V}\hat T e^{-\tau\hat V}$. In order to sample
expression (\ref{Eq-Worm-2}) an extended partition function is
considered by breaking up the propagator at imaginary time $\tau$ and
introducing a ``worm operator'' $\hat W=\sum_{ijkl}w_{ijkl}b_i^\dagger
b^{\phantom\dagger}_j f_k^\dagger f^{\phantom\dagger}_l$ that leads to $\mathcal
Z(\tau)=\textrm{Tr}e^{-(\beta-\tau)\hat H}\hat W e^{-\tau\hat H}$.
Complete sets of states are introduced between consecutive $\hat T$
operators to allow a mapping of the 1D quantum problem onto a 2D
classical problem where a standard Monte Carlo technique can be applied.
Measurements can be performed when configurations resulting in diagonal
matrix elements of $\hat W$ occur.  This way unphysical movements are
exploited to help explore the Hilbert space, but are ignored when
sampling for measurements.

As with pure bosonic systems, the evolution of the boson and fermion
densities $\rho_{B},\rho_{F}$ with the associated chemical potential
$\mu_{B},\mu_{F}$ identifies Mott insulating behavior \cite{fisher89}.  A
jump in $\mu$ signals a Mott phase where the compressibility
$\kappa_{B}=\partial \rho_B / \partial \mu_B$ or $\kappa_{F}=\partial
\rho_F / \partial \mu_F$ vanishes.

\begin{figure}[!htb]
\begin{center}
  \includegraphics[width=0.47\textwidth,angle=0]{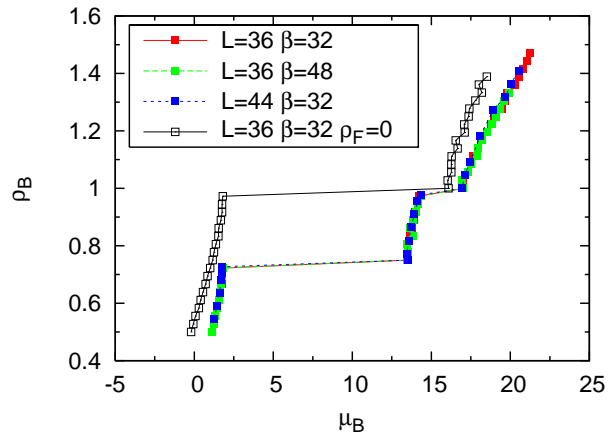}
\end{center}
\vspace{-0.5cm}
\caption{\label{Ubf16mu}
(Color online) $\rho_B$ as a function of chemical potential
$\mu_B$.  The fermion density is $\rho_F=1/4$ and the interaction
strengths are fixed at $U_{BB}=10$ and $U_{BF}=16$. There are Mott
plateaus at $\rho_B=1-\rho_F=3/4$ and $\rho_B=1$ as predicted by the
$t_B=t_F=0$ analysis.  The positions of the Mott lobes coincide for
different lattice sizes $L=36,44$ and temperatures $\beta=32,48$ to
within our error bars, which are smaller than the symbol size.  The
dependence of $\rho_B$ on $\mu_B$ in the absence of fermions is given
for comparison.}
\end{figure}

Quantities of interest that we measure include the bosonic superfluid
density and the fermionic stiffness,
\begin{eqnarray}
\rho^{\rm s}_B&=&\big\langle W_B^2\big\rangle L/2\beta
\nonumber \\
\rho^{\rm s}_F&=&\big\langle W_F^2\big\rangle L/2\beta \,\,.
\end{eqnarray}
Here $\langle W^2 \rangle$ are the associated winding numbers.
Correlations between the bosonic and fermionic winding numbers \cite{pollet06}
are determined by the combinations,
\begin{eqnarray}
\rho^{\rm s}_c&=&\big\langle (W_B+W_F)^2\big\rangle L/2\beta
\nonumber \\
\rho^{\rm s}_a&=&\big\langle(W_B-W_F)^2\big\rangle L/2\beta \,\,.
\end{eqnarray}
In addition to the usual
bosonic and fermionic Green function,
\begin{eqnarray}
G_{ij}^B&=&\big\langle b_i^\dagger b^{\phantom\dagger}_j\big\rangle
\nonumber \\
G_{ij}^F&=&\big\langle f_i^\dagger
f^{\phantom\dagger}_j\big\rangle \,\, ,
\end{eqnarray}
we also measure the composite anti-correlated two-body Green function
\begin{eqnarray}
G_{ij}^a&=&\big\langle b_i^\dagger b^{\phantom\dagger}_j f_j^\dagger f^{\phantom\dagger}_i\big\rangle  \,\,.
\end{eqnarray}
In $G_{ij}^a$, the fermion and boson propagate in
opposite directions
(one from $j$ to $i$ and one from $i$ to $j$).

The Fourier transforms of $G^B_{ij}$ and $G^F_{ij}$ give the densities
$n_B(k)$ and $n_F(k)$ in momentum space; $n_a(k)$ is
the Fourier transform of the composite two-body Green function 
$G^a_{ij}$.

We performed extensive checks of the code against other quantum Monte
Carlo simulations in the pure boson and pure fermion cases, and against
exact diagonalization and Lanczos calculations for mixed systems on
small lattices.

\begin{figure}[!htb]
\begin{center}
  \includegraphics[width=0.47\textwidth,angle=0]{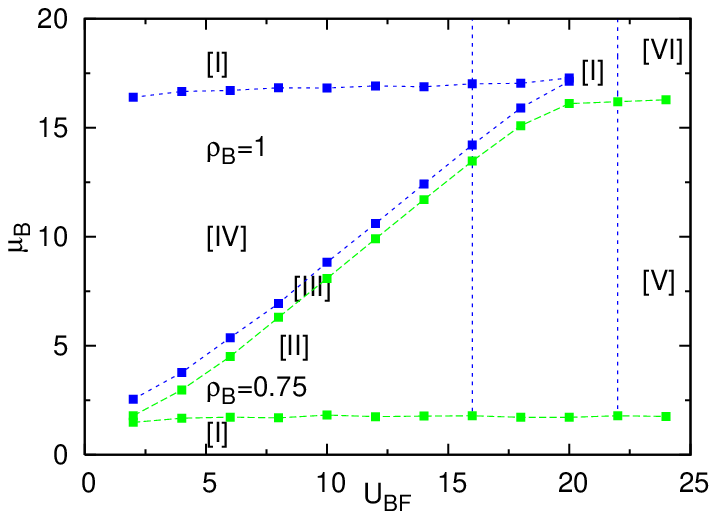}
  \includegraphics[width=0.47\textwidth,angle=0]{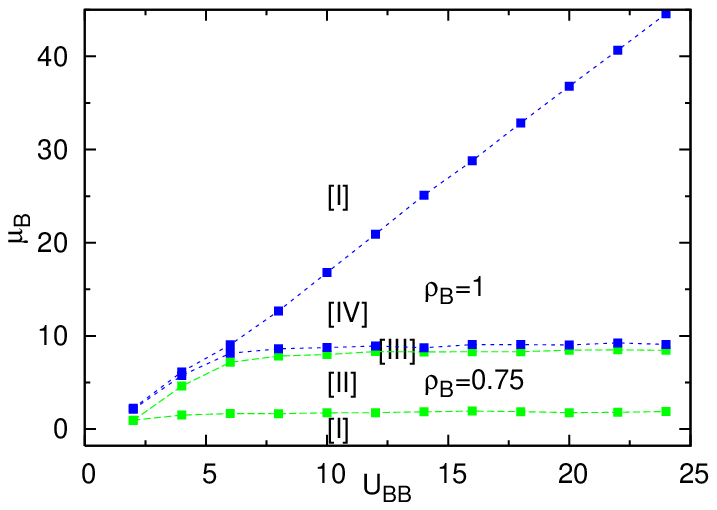}
\end{center}
\vspace{-0.5cm}
\caption{\label{phasediagram} (Color online) 
\underbar{Top panel:} 
Phase diagram in the
$\mu_B$-$U_{BF}$ plane for $U_{BB}=10$ and $\rho_F=1/4$ obtained by a
sequence of plots such as that in Fig.~\ref{Ubf16mu}. 
The vertical line at $U_{BF}=16$ corresponds to the coupling value in 
Fig.~\ref{Ubf16mu}.
Within the regions
labeled $\rho_B=0.75$ and $\rho_B=1$, the boson density is frozen even
though the chemical potential varies.  The $\rho_B=1$ lobe is pinched
off at $U_{BF} \approx 2U_{BB}$. The labeling of the phases is: {\bf I.}
Superfluid ($\rho^{\rm s}_B \neq 0$ $\rho^{\rm s}_F \neq 0$); {\bf II.}
Anti-Correlated phase ($\rho^{\rm s}_B \neq 0$, $\rho^{\rm s}_F \neq 0$,
$\rho^{\rm s}_B = \rho^{\rm s}_F$); {\bf III.} Anti-Correlated phase /
Relay Superfluid ($\rho^{\rm s}_B \neq 0$, $\rho^{\rm s}_F \neq 0$) (see
text for detailed explanation); {\bf IV.} Mott Insulator / Luttinger
liquid ($\rho^{\rm s}_B = 0$, $\rho^{\rm s}_F \neq 0$); {\bf V.}
Insulator, $\rho^{\rm s}_B = 0$ $\rho^{\rm s}_F = 0$;
{\bf VI.} Phase separation. \\
\underbar{Bottom panel:} 
Phase diagram in the
$\mu_B$-$U_{BB}$ plane for $U_{BF}=10$ and $\rho_F=1/4$.
The upper boundary of phase $II$ is defined by the increase of energy when 
a boson is added at filling $\rho_B + \rho_F=1$;
this is approximately $\mu_B = U_{BF}$ (for $U_{BF} < 2 U_{BB}$),
a line of slope 1 in the $\mu_B$-$U_{BF}$ plane
and slope 0 in the $\mu_B$-$U_{BB}$ plane.
A similar strong coupling analysis applies for the other 
boundaries.  (See also Fig.~\ref{phasediagcomp} and
[\onlinecite{lewenstein04}].)
}
\end{figure}

\section{Phase Diagram in the $\mu_B$-$U_{BF}$ plane}

We begin our determination of the phase diagram by calculating the
dependence of the density $\rho_B$ on chemical potential $\mu_B$,
mapping out the extent that the Mott plateaus described in the
introduction survive the introduction of quantum fluctuations $t_B,t_F$.
We examine a system with a fixed $U_{BB}=10$ and $\rho_F=1/4$ and focus
on the regions through $\rho_B \leq 3/2$ and $U_{BF} \leq 5U_{BB}/2$ in
the phase diagram.  The $t_B=t_F=0$ analysis suggests for $U_{BF} < 2
U_{BB}$ there will be plateaus with compressibility $\kappa=0$ at
$\rho_B+\rho_F=1$ (i.e.~$\rho_B=3/4$) caused by $U_{BF}$ and at
$\rho_B=1$ caused by $U_{BB}$.  Fig.~\ref{Ubf16mu} exhibits these
plateaus for $U_{BF}=16$ and $t_B=t_F=1$.  The complete phase diagram in
the $\mu_B$-$U_{BF}$ plane at fixed $U_{BB}=10$ is obtained by
replicating Fig.~\ref{Ubf16mu} for different $U_{BF}$, and is given in
Fig.~\ref{phasediagram}a.  For weak $U_{BF}$ the phase diagram is
dominated by the $\rho_B=1$ plateau where the chemical potential jumps by
$2U_{BB} - U_{BF} \approx 2U_{BB}= 20$.  As $U_{BF}$ increases, this
plateau shrinks and finally terminates at $U_{BF} \approx 2U_{BB} = 20$.
At the same time, the plateau at $\rho_B= 1 - \rho_F$ grows to
$U_{BF}=20$.  The explanation of the labeling of the different phases
({\bf I-VI}) will be given after we discuss the superfluid response of
the system.
Fig.~\ref{phasediagram}b shows the phase diagram in the 
$\mu_B$ and $U_{BB}$ plane.

\section{Superfluid Response at $\rho_B + \rho_F = 1$}

\begin{figure}[!htb]
\begin{center}
  \includegraphics[width=0.47\textwidth,angle=0]{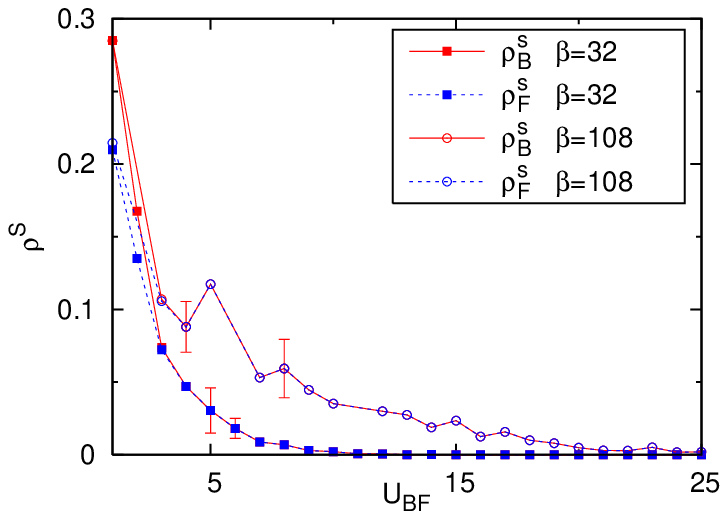}
  \includegraphics[width=0.47\textwidth,angle=0]{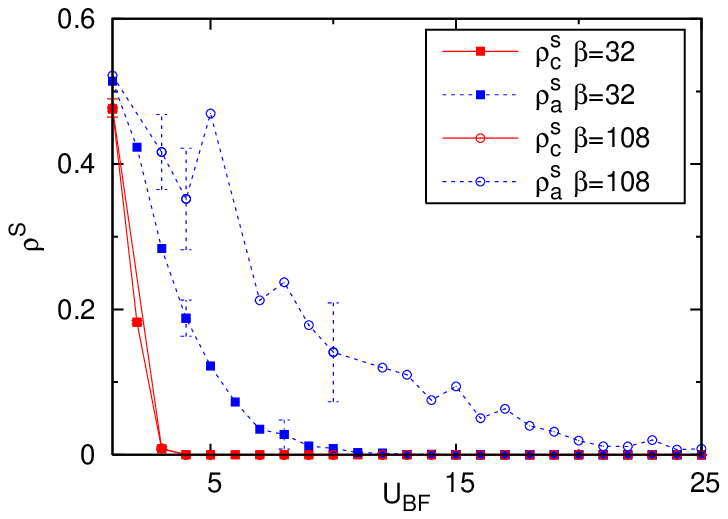}
\end{center}
\vspace{-0.5cm}
\caption{\label{WM1rhosbf0.75} 
(Color online) ``Horizontal''
sweep, that is fixed densities $\rho_B=3/4$ and $\rho_F=1/4$ and
varying $U_{BF}$, through the phase diagram of Fig.~\ref{phasediagram}.
As in Figs.~\ref{Ubf16mu} and \ref{phasediagram}, $U_{BB}=10$.
\underbar{Top panel:} Both the bosonic and fermionic species exhibit a
finite stiffness at weak coupling (region {\bf II} of phase diagram),
which decays as $U_{BF}$ increases until insulating behavior occurs
(region {\bf V} of phase diagram).  \underbar{Bottom panel:}  Near
$U_{BF}=0$ both $\rho^{\rm s}_c \neq 0$ and $\rho^{\rm s}_a \neq 0$ are
very similar.  However, quickly after turning on $U_{BF}$, the
correlated and anti-correlated stiffness show that the bosons and
fermions propagate in opposite directions $\rho^{\rm s}_a \neq  0$,
while $\rho^{\rm s}_c = 0$.}
\end{figure}

After determining the positions of the Mott plateaus, we examine the
stiffness and Green functions.  We take a ``horizontal'' cut through
Fig.~\ref{phasediagram}a by fixing $\rho_B+\rho_F=1$ ($\rho_B=3/4$) and
increasing $U_{BF}$.  In Fig.~\ref{WM1rhosbf0.75} we see that for
$U_{BF} \lesssim 2 U_{BB} = 20$ the interaction strength $U_{BF}$ is
small enough that fermions and bosons can briefly inhabit the same site.
Now, when a boson visits the site of a neighboring fermion (or vice
versa) it is equally likely that the fermion will exchange as for the
boson to return to its original site. Through these exchanges the bosons
and fermions can achieve anti-correlated winding around the lattice.
Thus, the bosonic superfluid density and the fermionic stiffness are
both non-zero and identical
\footnote {This phase has been termed ``Super-Mott" as a consequence of
its combining a non-zero gap with non-zero superflow \cite{rousseau01}.}.
However, as $U_{BF}$ increases past $U_{BF} \approx 2 U_{BB} = 20$ the
cost of double occupancy becomes prohibitive.  With its benefits
outweighed by energy penalties exacted by $U_{BF}$, all anti-correlated
``superfluidity" ceases. Pollet {\it et al.} \cite{pollet06} have argued
that this region exhibits phase separation.
Indeed we do detect a signal of phase separation 
through density structure factor. 
But the signal is weak, about 20 times weaker than what we get 
at phase {\bf VI} (next section),
and compressibility is close to zero,
so we label this region as an insulator. 

From the results depicted in Fig.~\ref{WM1rhosbf0.75} one should notice
that while for quantities like the energy and Mott gap $\beta=32$ is
sufficiently low for $L=32$ to capture the ground state behavior, for
stiffnesses one requires much lower temperatures.

\begin{figure}[!htb]
\begin{center}
  \includegraphics[width=0.47\textwidth,angle=0]{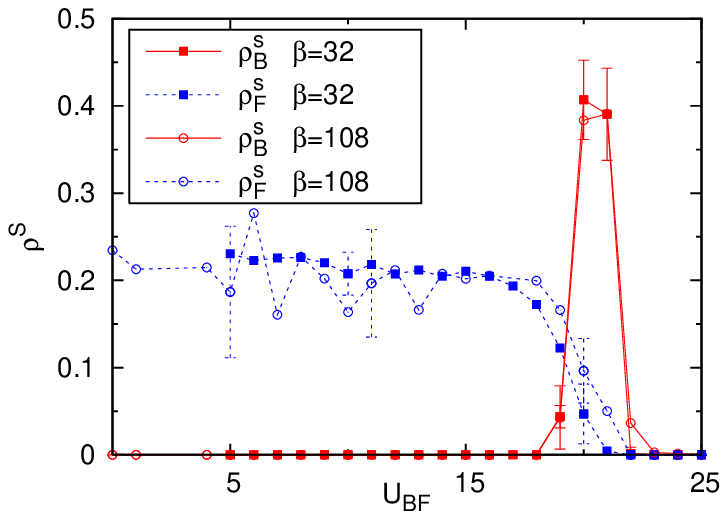}
  \includegraphics[width=0.47\textwidth,angle=0]{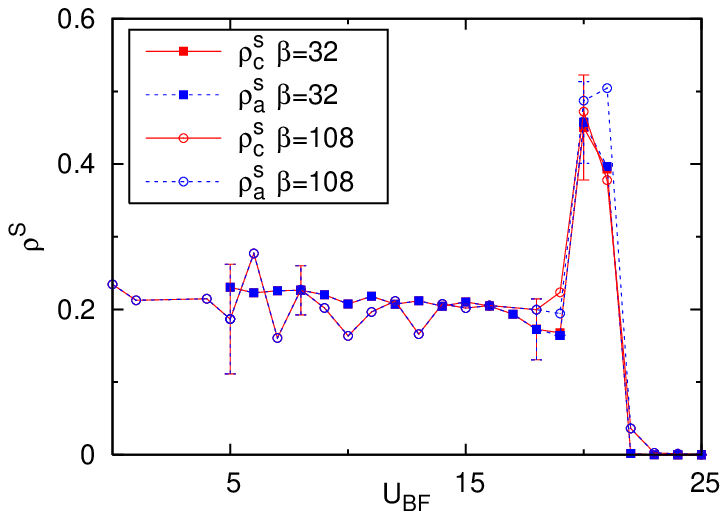}
\end{center}
\vspace{-0.5cm}
\caption{\label{WM1rhosbf1} (Color online)
Same ``horizontal cut'' as Fig.~\ref{WM1rhosbf0.75} except 
at commensurate density for the
boson species alone $\rho_B=1$ and $\rho_F=1/4$.
\underbar{Top panel:}
The bosons are
insulating at weak $U_{BF}$ within this Mott lobe of commensurate
bosonic filling (region {\bf IV} of phase diagram).
However the fermions are free to flow on the uniform
boson background and have nonzero stiffness.
Upon emerging from the lobe, at $U_{BF}\approx 2 U_{BB}$,
$\rho^{\rm s}_B$ becomes nonzero in a window where the two repulsions
work against each other.
After the peak, the system becomes phase separated and $\rho^{\rm s}_B$ and $\rho^{\rm s}_F$ go to zero.
\underbar{Bottom panel:}
The correlated and anti-correlated stiffnesses are essentially
equal throughout the weak coupling because the flow is
dominated by fermions.  However, in the window the
anti-correlated stiffness increases beyond the correlated stiffness
in a weak simulacrum of phase {\bf II} (as discussed in the text).}
\end{figure}

\section{Superfluid Response at $\rho_B = 1$}

Although it shares the property that $\kappa_B=0$ with the
$\rho_B+\rho_F=1$ lobe, a `horizontal' cut (Fig.~4) 
through the $\rho_B=1$ Mott
lobe exhibits rather different superfluid response.  This trajectory
initially lies within the Mott lobe and then emerges into a region of
non-zero compressibilities.  As expected, the plateau in $\rho_B$ (Mott gap)
indicates the bosons are locked into place by the strong $U_{BB}$, and
as a consequence $\rho^{\rm s}_B = 0$ (Fig. \ref{WM1rhosbf1}).
Throughout this boson Mott lobe the fermions are, however, free to slide
over the bosons and so $\rho^{\rm s}_F$ is non-zero.  In this region, as
expected, the fermion compressibility $\kappa_F$ is nonzero.

The Bose-Fermi repulsion $U_{BF}$ competes with $U_{BB}$ and, in a
window around $U_{BF} \approx 2U_{BB}$, it is energetically equivalent
for a boson to share a site with another boson as with a fermion.  The
Mott lobe is terminated and a superfluid window opens for both species.
Finally, for $U_{BF} > 2 U_{BB}$, it is energetically unfavorable for
a boson to share a site with a fermion.
We enter a region of phase separation where 
superflow for both species stops, but the compressibilities $\kappa_B$ and $\kappa_F$ are nonzero.
We also confirm phase separation through a density structure factor.
See also \cite{hebert07,mering08}.

\section{Superfluid Response At General Filling}

\begin{figure}[!htb]
\begin{center}
  \includegraphics[width=0.47\textwidth,angle=0]{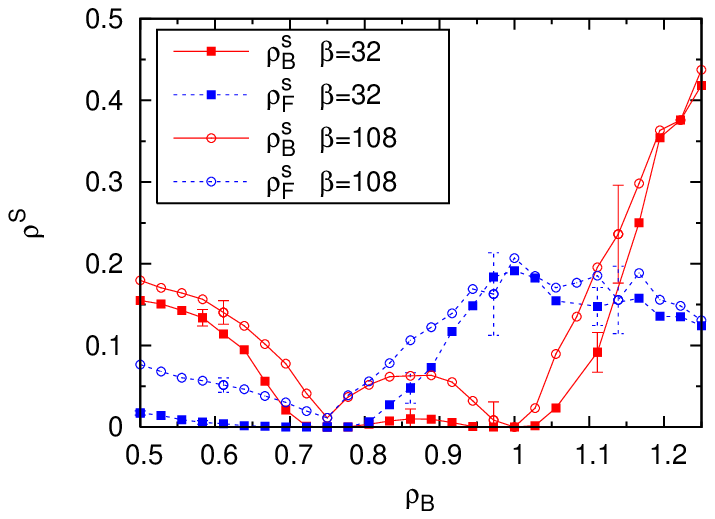}
  \includegraphics[width=0.47\textwidth,angle=0]{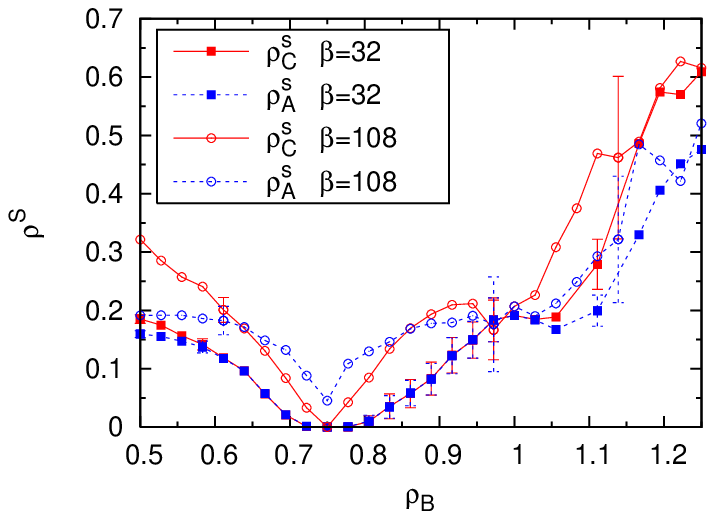}
\end{center}
\vspace{-0.5cm}
\caption{\label{Ubf16rhosf} (Color online) ``Vertical" sweep across $\rho_B$ at $\rho_F=1/4$,
$U_{BB}=10$ and $U_{BF}=16$.
\underbar{Top panel:} The boson superfluid density changes much less
as $\beta$ is increased from $\beta=32$ to $\beta=108$ in the
superfluid phase {\bf I} at $\rho_B < 3/4$ and $\rho_B>1$ than
for phase {\bf III} $3/4 < \rho_B < 1$.  This is a
hallmark of the ``relay'' superfluid discussed in the text.
\underbar{Bottom panel:} The correlated winding decreases to zero at $\rho_B = 3/4$
while the anti-correlated winding remains finite.  As $\rho_B$ is increased
beyond $3/4$ the correlated winding increases and overtakes the anti-correlated
winding.  This is another sign of the ``relay'' superfluid.}
\end{figure}

\begin{figure}[!htb]
\begin{center}
  \includegraphics[width=0.47\textwidth,angle=0]{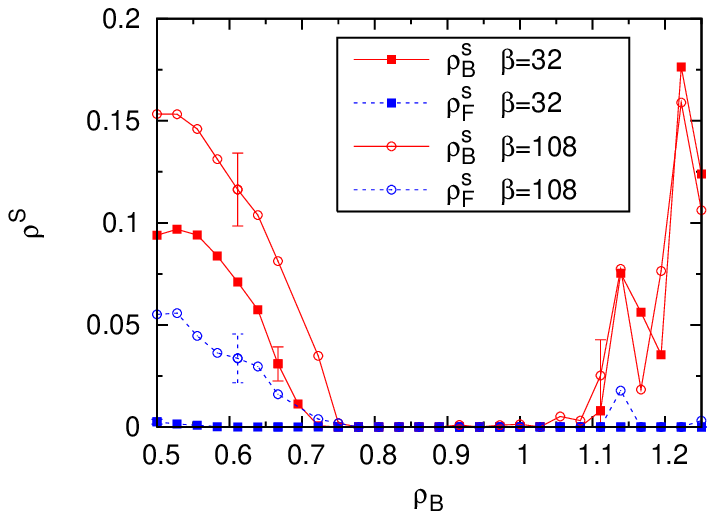}
  \includegraphics[width=0.47\textwidth,angle=0]{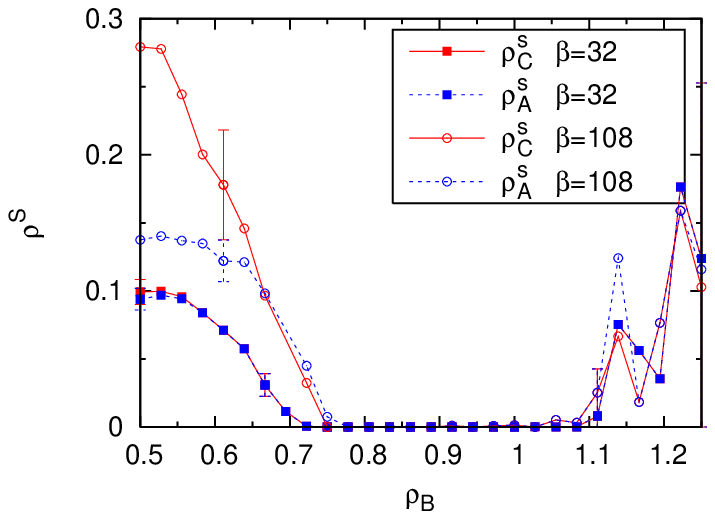}
\end{center}
\vspace{-0.5cm}
\caption{\label{Ubf24rhosf} (Color online) ``Vertical" sweep across $\rho_B$ at $\rho_F=1/4$,
$U_{BB}=10$ and $U_{BF}=24$. Unlike the case for the weaker coupling
$U_{BF}=16$ in Fig.~ \ref{Ubf16rhosf}, both superfluid densities vanish
in the insulating region of the Mott lobe at commensurate
total filling.}
\end{figure}

Further insight into the physics of this phase diagram can be obtained
by measuring the superfluid response along the same `vertical' cuts
through the phase diagram as done in Figs.~\ref{Ubf16mu} and
\ref{phasediagram}, in which $\rho_B$ is varied at fixed $U_{BF}$. In
Figs.~\ref{Ubf16rhosf} and \ref{Ubf24rhosf}, we show the result.
Distinctive densities in the latter figures are $\rho_B=3/4$ (so
that $\rho_B + \rho_F \; = \; 1$) and $\rho_B=1$. We discuss first
(Fig.~\ref{Ubf16rhosf}) the case of $U_{BF}=16$, where increasing
$\rho_B$ cuts through both Mott lobes.  The bosonic superfluid density
vanishes at $\rho_B=1$, dips at $\rho_B+\rho_F=1$, and is nonzero above,
below, and between the lobes.  The fermion superfluid density is never
driven to zero in this cut, and only dips at the special value
$\rho_B=3/4$ where the commensurate total density works against
superfluidity. In the case of $U_{BF}=24$, Fig.~\ref{Ubf24rhosf}, as
$\rho_B$ increases we cut through only the $\rho_B=3/4$ lobe.  Here the
superfluid density is pushed to zero for the entire region between
$\rho_B=3/4$ and $\rho_B=1$, and is non-zero without.

We now fill in the labeling of the phase diagram of
Fig.~\ref{phasediagram}. 
A gapless superfluid phase ({\bf I}) with ($\rho^{\rm
s}_B \neq 0$ and $\rho^{\rm s}_F \neq 0$) exists at low filling of the
lattice $\rho_B+\rho_F < 1$.  
When the combined filling of the two
species becomes commensurate, an anti-correlated ({\bf II}) phase
appears in which $\rho^{\rm s}_B \neq 0$ and $\rho^{\rm s}_F \neq 0$,
but $\rho^{\rm s}_B = \rho^{\rm s}_F$.  This phase is characterized by
superflow of the two species in opposite directions and is gapped to
the addition of bosons or fermions.
The usual bosonic
Mott insulator, phase {\bf IV}, occurs at commensurate boson densities.
However, it can be melted by increasing $U_{BF}$ since the jump in
bosonic chemical potential (Mott gap) is reduced to $2U_{BB}-U_{BF}$.
There is no jump in $\mu_F$.
Eventually quantum fluctuations break this gap and superflow is allowed.
When $U_{BF}$ exceeds $2U_{BB}$, all superflow stops and we enter the
insulating region {\bf V} of the phase diagram.

We speculate that the nature of the superfluidity in the narrow phase
{\bf III}, which exists between the two Mott lobes is an unusual
``relay" process.  It is similar to the usual superfluid which exists
between Mott lobes in the single species model, in that $\rho^{\rm s}_B
\neq 0$.  However, the temperature scale at which superfluid
correlations build up is dramatically reduced.  This occurs because the
bosons can exhibit superflow only by traveling along with a fermion
partner, and being handed off from fermion to fermion in order to wind
around the entire lattice.  The point is that because $2 U_{BB}$ exceeds
$U_{BF}$ the bosons doped into the lattice above $\rho_B=1-\rho_F=3/4$ are
forced to sit on a fermion.  They cannot hop off, but the fermion can
move since it has already paid $U_{BF}$ to share a site with a boson.
Now, the fermions cannot pass each other once a fermion riding atop
bosons runs into a fermion alone on a site.  The fermion without a boson
cannot move out of the other fermion's way either.  However, the boson
sharing a site with the mobile fermion can then hop to the immobile
fermion at no energy cost.  Thus, the boson is passed from one fermion
to the other, granting it mobility.  Signatures of this phase are the
lower value of the temperature at which the superfluid density builds
up, that $\rho^{\rm s}_F > \rho^{\rm s}_B$, and more correlated
winding than anti-correlated.  However, there is nothing preventing lone
fermions from acting as in the anti-correlated superfluid phase.
Unfortunately this means that potential signals are masked.  While we
do see some of these signatures (Fig.~\ref{Ubf16rhosf}) in the specified
region, the numbers are not completely conclusive and will require
further investigation.

\section{Momentum distribution functions}

\begin{figure}[!htb]
\begin{center}
  \includegraphics[width=0.47\textwidth,angle=0]{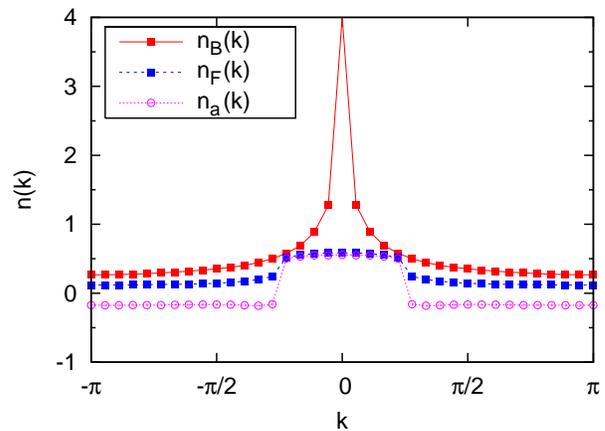}
\end{center}
\vspace{-0.5cm}
\caption{\label{k1} (Color online) {\bf I.} $U_{BF}=16$,
$U_{BB}=10$, $\beta=108$ and $N_{B}=20$;
Momentum distributions for bosons and fermions, and Fourier transform of 
the anti-correlated pairing 
($n_a(k)$) 
Green function. The sharp peak in 
bosonic momentum distribution indicates the presence of a quasi-condensate, 
while fermions have a plateau indicating Luttinger liquid like behavior with 
a clear Fermi momentum, a property that is also shared by the composite
fermions described by the anti-correlated pairing.}
\end{figure}

\begin{figure}[!htb]
\begin{center}
  \includegraphics[width=0.47\textwidth,angle=0]{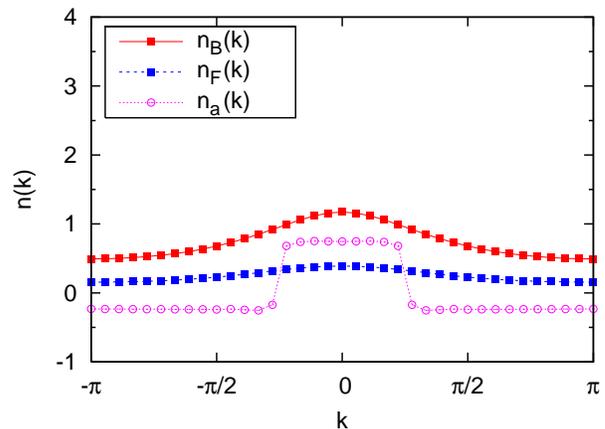}
\end{center}
\vspace{-0.5cm}
\caption{\label{k2} (Color online) {\bf II.} $U_{BF}=16$,
$U_{BB}=10$, $\beta=108$ and $N_{B}=27$.  Bosons do not have a peak at
$k=0$ and the Fermi momentum is washed out, both reflecting the onset of
short range one-particle correlations. On the other hand, the plateau in
the Fourier transform of the anticorrelated pairing shows that the
composite fermions formed by pairing a fermion and a boson have a
well defined Fermi momentum \cite{pollet06}.}
\end{figure}

\begin{figure}[!htb]
\begin{center}
  \includegraphics[width=0.47\textwidth,angle=0]{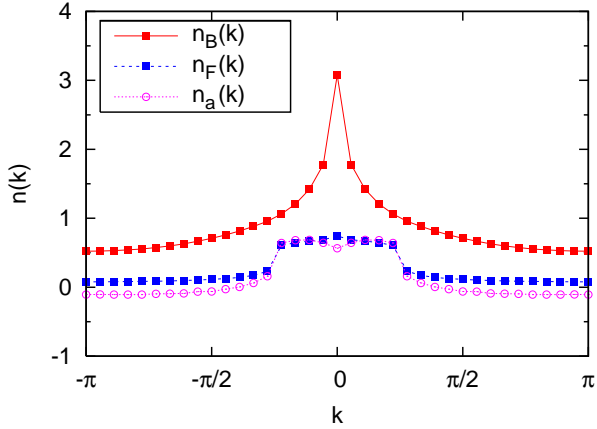}
\end{center}
\vspace{-0.5cm}
\caption{\label{k3} (Color online) {\bf III.} $U_{BF}=16$,
$U_{BB}=10$, $\beta=108$ and $N_{B}=32$.
Qualitatively, this picture is similar to Fig.~\ref{k1} - we have a peak
in the bosonic momentum distribution and a plateau in the fermionic
and anticorrelated pairing momentum distributions, all indicating 
power-law decaying correlations of their corresponding real space 
Green functions.}
\end{figure}

\begin{figure}[!htb]
\begin{center}
  \includegraphics[width=0.47\textwidth,angle=0]{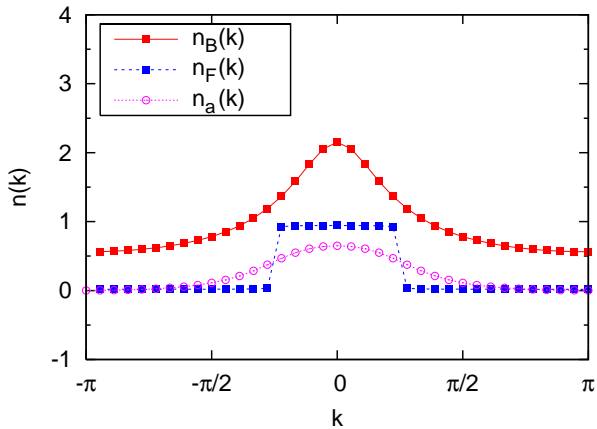}
\end{center}
\vspace{-0.5cm}
\caption{\label{k4} (Color online) {\bf IV.} $U_{BF}=16$,
$U_{BB}=10$, $\beta=108$ and $N_{B}=36$. 
There is no sharp peak in $n_B(k)$ and a plateau in $n_F(k)$
is present.   This phase is a 
Mott Insulator for bosons and Luttinger liquid behavior for the fermions.
In this case the composite fermions do not exhibit a Fermi momentum.}
\end{figure}

\begin{figure}[!htb]
\begin{center}
  \includegraphics[width=0.47\textwidth,angle=0]{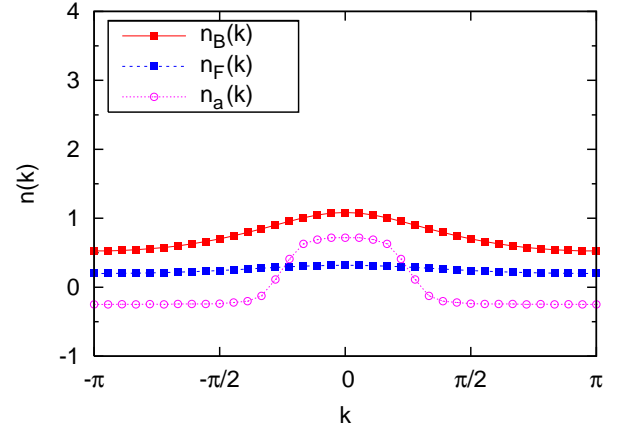}
\end{center}
\vspace{-0.5cm}
\caption{\label{k6} (Color online) {\bf V.} $U_{BF}=30$,
$U_{BB}=10$, $\beta=108$ and $N_{B}=27$. 
The momentum distribution functions in this case exhibit the behavior expected from
an insulator, i.e., no sharp peak in $n_B(k)$, no plateau in $n_F(k)$, and
no Fermi edge in $n_a(k)$.}
\end{figure}

\begin{figure}[!htb]
\begin{center}
  \includegraphics[width=0.47\textwidth,angle=0]{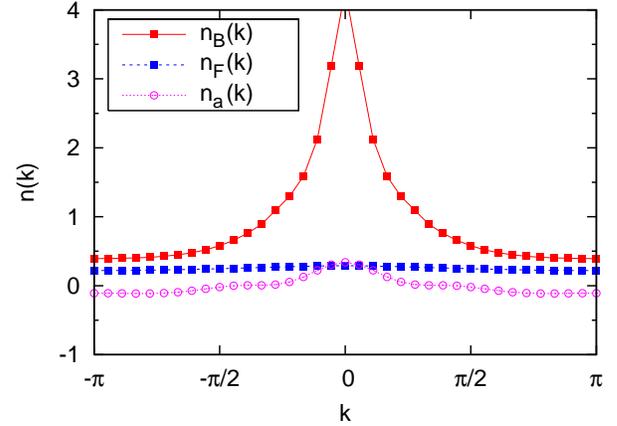}
\end{center}
\vspace{-0.5cm}
\caption{\label{k7} (Color online) {\bf VI.} $U_{BF}=28$,
$U_{BB}=10$, $\beta=108$ and $N_{B}=36$. 
Phase separation. 
The boson momentum distribution function has a peak indicating that there may be 
a kind of superflow in their separate area.
Fermions, on the other hand, behave as an isolator.
The $n_a(k)$ curve indicates that the coupling between bosons and fermions is weak, as we would expect in phase separation. 
}
\end{figure}

To further explore the nature of the phases we turn to the
momentum distributions for the bosons, fermions, and
anti-correlated pairing - Fig.~\ref{k1} - \ref{k7}.  Each plot is made
at $\beta=108$ and correspond to the parameter choices:
{\bf I.} $U_{BF}=16$ and $N_{B}=20$; 
{\bf II.} $U_{BF}=16$ and $N_{B}=27$; 
{\bf III.} $U_{BF}=16$ and $N_{B}=32$;
{\bf IV.} $U_{BF}=16$ and $N_{B}=36$;
{\bf V.} $U_{BF}=30$ and $N_{B}=27$;
{\bf VI.} $U_{BF}=28$ and $N_{B}=36$.  
In the superfluid phase ({\bf I.}), there is 
a peak in the boson distribution and a plateau in the
fermion distribution, implying quasi-condensation in the bosonic sector
and Luttinger liquid-like behavior in the fermionic one. In the
anti-correlated phase ({\bf II.}) there is neither of the former
behaviors, but the Fourier transform of the anti-correlated pairing
Green function has a clear Fermi momentum showing Luttinger-like
physics of the composite fermions (formed by pairing a fermions and a
boson) \cite{pollet06,foot}.  The ``relay'' superfluid phase ({\bf III.})
displays momentum distributions that are similar to the ones of
superfluid phase ({\bf I}).  Next, in the Mott insulator / Luttinger liquid
phase ({\bf IV.}) one can see a clear Fermi momentum in the bare
fermion $n_F(k)$ and a very smooth behavior of $n_B(k)$ and $n_a(k)$,
which show that their real space Green function counterparts are
decaying exponentially. In the insulating phase ({\bf V.}) all
the correlations decay exponentially and their corresponding
momentum distribution functions are smooth functions of $k$.
In the case of phase separation ({\bf VI.})
the bosonic
momentum distribution is similar to the superfluid,
while fermionic distribution is insulating.
 
\section{Connection to previous theoretical work}

\begin{figure}[!htb]
\begin{center}
  \includegraphics[width=0.47\textwidth,angle=0]{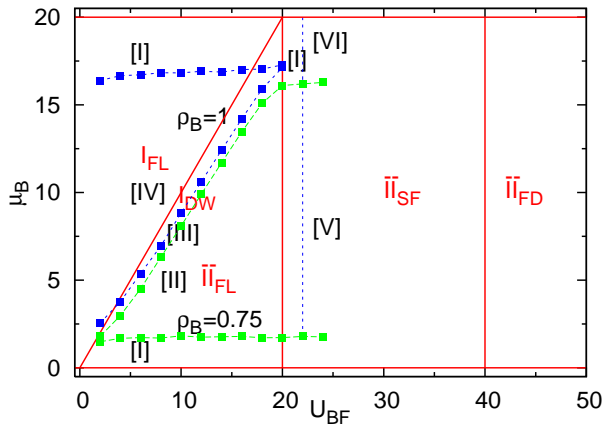}
\end{center}
\vspace{-0.5cm}
\caption{\label{phasediagramL} (Color online) 
A comparison of our 
phase diagram (Fig.~\ref{phasediagram}) with the strong coupling
boundaries.
Symbols and dashed lines are the results of the present QMC work, while
the solid lines are for $t_B=t_F=0$.
Unsubscripted Roman symbols denote our phases while
subscripted Roman symbols are the labeling of Lewenstein
{\it et al.} \cite{lewenstein04}.
\label{phasediagcomp}
}
\end{figure}

As reviewed in the introduction, there is an extensive theoretical
literature on Bose-Fermi mixtures.  We now make more detailed contact
with previous work, first by comparing our results to the strong coupling 
phase diagram of Lewenstein {\it et al.} (LSBF) \cite{lewenstein04}.  
Fig.~\ref{phasediagramL} combines our results and those of
LSBF.  Besides the quantitative agreement, we note the
following correspondences:  LSBF's region
$0\le \bar\mu \le 1$ is analogous to our $0\le \mu \le 20$, and $0\le
\alpha \le 1$ to our $0\le U_{BF} \le 20$.  
Furthermore,
our phase $II$ (Anti-Correlated phase) corresponds to LSBF's phase
$\bar{II}_{FL}$ (Fermi liquid of composite fermions formed by one bare
fermion and bosonic hole); our phase $IV$ (Mott insulator / Luttinger
liquid) to LSBF's phase ${I}_{FL}$ (Fermi liquid); and finally
our phase $III$ (Anti-Correlated phase / Relay superfluid) to
LSBF's phase ${I}_{DW}$ (Density wave phase). These three phases
have similar qualities and occur approximately at the same locations at
our and LSBF's phase diagrams.

Both calculations suggest the existence of composite particles.  Our
phase $V$ (Insulator) corresponds to LSBF's phase $\bar{II}_{FD}$,
a region of fermionic domains of composite fermions formed by one bare
fermion and bosonic hole.  There is one case when the phases do not seem
to correspond well, namely LSBF's phase $\bar{II}_{SF}$ which is a
superfluid of composite fermions formed by one bare fermion and bosonic
hole.  Our results (Fig.~3) instead suggest that in this region of
$U_{BF} > 2 U_{BB}$, the superfluid densities vanish, or are very
small.

\section{Experimental issues}

Albus {\it et al.} \cite{albus03} have given the correspondence between
Hubbard model parameters $U_{BB}, U_{BF}, t_B, t_F$ and experimentally
controlled parameters.  $U_{BB}$ and $U_{BF}$ are determined by the
optical lattice depth, laser wavelength, and harmonic oscillator
lengths, as well as by the 
scattering lengths $a_{BB}$ and $a_{BF}$ which 
can be tuned by traversing a Feshbach resonance.  Similarly,
the hoppings $t_B$ and $t_F$ follow from the lattice depth 
and atomic masses.
It is possible to choose experimentally reasonable values of 
these parameters to correspond to the energy scales chosen in
our paper.  For example, 
following Albus {\it et al},
for a $^{87}Rb, ^{40}K$ mixture and
laser wavelength $600 nm$, 
$a_{BB}=100 a_0$, $a_{BF}=123.74 a_0$,
and $V_0=0.7614$ in units of boson recoil energy,
with $l_B^\perp= 17.04nm$,
we get in units of $t_B$: 
$t_B=1$,
$t_F=2$, and
$U_{BB}=U_{BF}=10$.
In this paper we have used $t_B=t_F=1$, which would be accessible in
mixtures with $m_F \approx m_B$ such as $^{40}K, ^{41}K$.\\

\section{Conclusions}

In conclusion, we have mapped out the boson density - interaction
strength phase diagram of Bose-Fermi mixtures.  The Mott lobe at
commensurate total density has nontrivial superfluid properties, where
the two components of superflow can be nonzero and anti-correlated, or
both vanish. Likewise the Mott lobe at commensurate bosonic density has
vanishing boson superflow and nonzero fermion stiffness.  $\rho^{\rm
s}_B$ is nonzero upon emerging from this lobe where the balance between
boson-boson and boson-fermion repulsions opens a superfluid window, with
anti-correlated superflow.  The superfluidity between the
two Mott regions may be of a novel type where the bosons travel along
with the fermions (chosen to have relatively low density in this work).
As a consequence, the superfluid onset temperature is significantly
reduced.
Finally, we have discussed the signatures of the above phases in the
momentum distribution function of fermions and bosons, which can be
measured in time of flight experiments.

We acknowledge support from the National Science Foundation Grant No.\
ITR-0313390, Department of Energy Grant No.\ DOE-BES DE-FG02-06ER46319,
and useful conversations with G. G. Batrouni and T. Byrds.  This work is
part of the research program of the Stichting voor Fundamenteel
Onderzoek der materie (FOM), which is financially supported by the
Nederlandse Organisatie voor Wetenschappelijk Onderzoek (NWO).


\begin{thebibliography}{10}
\bibitem{foot}
Unlike $n_B(k) = \langle\psi_0 | b^\dag(k) b(k) |\psi_0\rangle$, which
is the length of a vector and hence must be positive, there is no such
constraint on $n_a(k)$, which is the Fourier transform of $G_{ij}^a =
\big\langle b_i^\dagger b^{\phantom\dagger}_j f_j^\dagger
f^{\phantom\dagger}_i\big\rangle$.
We have verified that the same small negative values of $n_a(k)$ in the
figures are also obtained in exact diagonalization on small clusters.
\end{thebibliography}
\end{document}